# Equation of state for hard sphere fluids offering accurate virial coefficients

Jianxiang Tian[1,3,5], Hua Jiang[2], Yuanxing Gui[3] and A. Mulero[4]

[1]Shandong Provincial Key Laboratory of Laser Polarization and Information Technology Department of Physics, Qufu Normal University, Qufu, 273165, P. R. China

[2]Department of Modern Physics, University of Science and Technology of China, Hefei, 230026, P. R. China

[3]Department of Physics, Dalian University of Technology, Dalian, 116024, P. R. China

[4]Department of Applied Physics, University of Extremadura, Badajoz 06072, Spain

[5]Corresponding author, Email address: jxtian@dlut.edu.cn;

**ABSTRACT:**

The asymptotic expansion method is extended by using currently available accurate values for the first ten virial coefficients for hard sphere fluids. It is then used to yield an equation of state for hard sphere fluids, which accurately represents the currently accepted values for the first sixteen virial coefficients and compressibility factor data in both the stable and the metastable regions of the phase diagram.

# Introduction

As is well known, the hard sphere (HS) fluid is defined by an interaction potential that considers only the repulsive forces among molecules. The simplicity of this model allows one to calculate its thermodynamic properties by obtaining analytical solutions for certain theories or by performing computer simulations. As the structure of real fluids is mainly determined by repulsive forces, the HS model is the simplest and most widely used model to describe the behaviour of fluids [1]. In particular, it plays an important role in perturbation theories [2], in statistical associating fluid theories [3], *etc*. Moreover, it has served as the basis for the advance of science in the fields of general liquids, amorphous solids, liquid crystals, colloids, granular matter, *etc*.[1]

In order to describe the thermodynamic properties of fluids, the equation of state (EOS) is the most important relationship that one requires. Unfortunately, there is no exact theoretical solution for the EOS of HS systems (except for the one-dimensional case). As a consequence, a great variety of expressions for the HS EOS can be found in the literature. An extensive review including more than eighty analytical expressions for the HS EOS has been published recently[4]. Most expressions were obtained from knowledge of the virial coefficients and/or by directly fitting computer simulation data. There has been major progress in the reproduction of computer simulation data in the stable region of the phase diagram [4], and some recently proposed expressions can reproduce them with extremely high accuracy. Unfortunately, there is less accuracy in reproducing either the highest known virial coefficients or the metastable region.

Virial coefficients can be considered as the cornerstones of the theory of fluids at low and medium densities [5], and, as is well-known, they are the coefficients in the density expansion of the EOS expressed via the compressibility factor, $Z$, as follows:

$$Z = \frac{P}{\rho k_B T} = 1 + \sum_{i=2}^{\infty} B_i y^{i-1} \qquad (1)$$

where $P$ is the pressure, $\rho$ the density, $T$ the temperature, and $k_B$ Boltzmann's constant. The packing fraction, $y$, is defined as the ratio between the volume occupied by the particles and the total volume. In HS fluids, $y = \pi\rho/6$.

The virial coefficients $B_i$ are defined by exact formulas in terms of integrals whose integrands are products of Mayer functions. The problem of obtaining an EOS for the fluid could be solved if one could determine all the coefficients in the infinite virial expansion. In particular, for hard spheres the integrals are numbers (they do not depend on temperature), but unfortunately only the first four can be calculated analytically [6]:

$$B_2 = 4, \quad B_3 = 10, \quad B_4 = \frac{2707\pi + [438\sqrt{2} - 4131\arccos(1/3)]}{70\pi} = 18.364768....$$

The higher virial coefficients must be calculated numerically [6-8]. We consider here the values for the fifth to tenth virial coefficients obtained by Clisby and McCoy [9-10] as $B_5 = 28.224512$, $B_6 = 39.815148$, $B_7 = 53.344420$, $B_8 = 68.537549$, $B_9 = 85.812838$, and $B_{10} = 105.775104$. Predicted values [9-10] for $B_{11}$ to $B_{16}$ are listed in Table 2.

As indicated above, there have been several dozens of equations for the hard-sphere fluid developed by different methods [1] ranging from statistical mechanics, such as the scaled particle theory [6], the integral equation theory [11-12], and the exclusion factor theory [13], to Padé aproximants[14]. Most of the available EOSs for the HS fluid accurately reproduce the first virial coefficients and the computer simulation data for the compressibility factor in the stable region of the phase diagram ($y < 0.494$). For instance, the most popular equation, the Carnahan-Starling expression (CS)[15], accurately represents the lower virial coefficients to $B_3$ as

$$Z_{CS} = \frac{1+y+y^2-y^3}{(1-y)^3} = 1+4y+10y^2+18y^3+o(y^4) \tag{2a}$$

Moreover, it gives good results at low densities. Unfortunately, no higher virial coefficients can be accurately reproduced with this expression, and no adequate results are obtained in the metastable density range. [1, 16]

Several other analytical expressions have been proposed for the HS EOS improving CS and reproducing the first seven or eight virial coefficients [14,17-22], but they use older values for those coefficients and do not give good results for the compressibility factor at high densities [1]. In fact, most of those EOSs are simple or even very simple expressions with a low number of parameters, and most of them were constructed mainly to be used as reference part in a perturbation scheme in which the effect of attractive forces is subsequently added. In these cases[22], the accuracy of the HS EOSs is sacrificed in order to have a simple complete expression and adequate results for more real fluids. Clear examples are the EOSs proposed by Yelash and Kraska[22], which were constructed as the simplest possible in order to analyse the liquid-liquid closed loop behaviour, and where the position of the pole had to be properly chosen.

When the previously mentioned EOSs were proposed, neither very accurate values for the compressibility factor at high densities nor the higher virial coefficients were available. Subsequently, Kolafa et al.[23] obtained highly optimized molecular dynamics computer simulation calculations in the range of reduced densities 0.20-1.03. Following the idea of Barboy and Gelbart [17], and with the aim of leading to good results in both the stable and the metastable regions, their data were fitted to power series in $y/(1-y)$, as follows:

$$Z_{KLM} = \sum_i C_i \left(\frac{y}{1-y}\right)^i \tag{3}$$

where $C_i$ are coefficients to be determined. The first six were determined to reproduce the first six virial coefficients (they specifically recalculated the fifth and sixth virial coefficients), whereas the others were considered to be adjustable parameters. As was noted by those authors, some of these coefficients can be zero (the coefficients are given in Refs. 1 and 23). Two different expressions were proposed: one considering only the region $\rho \leq 0.98$ (referred as KLM1) and the other for $\rho \leq 1.03$ (referred as KLM2).

As said before, Clisby and McCoy [9-10] have recently evaluated the first ten virial coefficients for hard spheres in dimensions from 2 to 8. This allowed them to propose the following two Padé approximants[10]

$$Z_{CM1} = \frac{1 + 2.03995984\eta + 3.3399824\eta^2 + 2.332827008\eta^3 + 0.900518016\eta^4}{1 - 1.96004012\eta + 1.180142976\eta^2 - 1.152111936\eta^3 + 1.4787069952\eta^4 - 0.5607328768\eta^5} \tag{4}$$

$$Z_{CM2} = \frac{1 + 3.21634468\eta + 7.4659272\eta^2 + 9.0866976\eta^3 + 7.613060608\eta^4 + 2.8869096448\eta^5}{1 - 0.78365528\eta + 0.600548464\eta^2 - 3.843711104\eta^3 + 3.149556992\eta^4}$$

(5)

Accordingly to the authors, these expressions are valid only at low densities.

Recently, Liu[16] developed an analytical equation of state for the entire stable and metastable regions. He used a potential energy landscape analysis to derive the following expression:

$$Z_L = Z_{L0} + \frac{0.31416\eta}{1 - 1.573357\eta} + 4.1637 \cdot 10^{10} \eta^{40} - 2.3452 \cdot 10^{11} \eta^{42} + 3.6684 \cdot 10^{11} \eta^{44}$$

(6)

where $Z_{L0}$ is defined as

$$Z_{L0} = 1 + \frac{3.68584\eta}{1 - 2.5848\eta + 1.9499\eta^2 - 0.172284\eta^3 - 0.16012\eta^4} \quad (7)$$

and was constructed by Liu by taking into account the values of the first twelve virial coefficients as published by Clisby and McCoy[9-10]. In Eq. (6) the coefficients were obtained through a fitting procedure to computer simulation data over the entire density range.

Recently, Khanpour and Parsafar[24] have proposed the asymptotic expansion method as a simple way to generate various EOSs in a unifying way which is also valid for the two-dimensional system[25]. In this method, the accurate virial coefficients are used as reference values to construct the EOSs. In particular, for the HS case they developed several EOSs by using the values of the first four virial coefficients. The proposed EOSs reproduce the computer simulation data for the compressibility factor with moderate accuracy at intermediate densities, but they cannot be applied to the high-density or metastable ranges.

Very recently, Santos and López de Haro [26] proposed a branch-point approximant for HS EOS, which reads

$$Z_{SH} = 1 + \frac{1 + c_1 y + c_2 y^2 + c_3 y^3 - (1 + 2a_1 y + a_2 y^2)^{3/2}}{A(1-y)^3} \qquad (8)$$

Where $A$, $a_1$, $a_2$, $c_1$, $c_2$, $c_3$ are parameters determined by known first seven virial coefficients. The authors denote the proposed equation gives out satisfactory prediction to higher virials and is in good agreement with simulation data.

In this paper, we extend the asymptotic expansion method of Khanpour and Parsafar[24] to find new, accurate equations of state for hard spheres. In the following section, the asymptotic expansion method is explained and used to generate new EOSs. Then some constraints are considered in order to choose the most appropriate expression. In the Results section, the results obtained from the proposed EOSs are compared with the accurate data for the virial coefficients and compressibility factor. Finally, the conclusions are summarized.

## The New Equation of State

In accordance with the aforementioned asymptotic expansion method (AEM), we assume that the hard-sphere equation of state can be written as:

$$z_{AEM} = \sum_{k=i}^{j} a_k x^k \qquad j > i, \quad i, j, k \in N \qquad (9)$$

where $a_k$ are coefficients to be determined, $x = 1/(y-b)$, $b$ being the radius of convergence of the virial expansion. Thus, for instance, with $b = 1$ and taking the integer values of the first three virial coefficients, one obtains the CS EOS as

$$Z_{CS} = 1 + 2x_1 - 2x_1^3 \qquad (2b)$$

where $x_1 = 1/(y-1)$.

With Eq. (9) defined, the following step is to consider some constraints in order to select the appropriate number of coefficients and convergence radius. These constraints are: **(i)** consistency between the calculated accurate virial coefficients and the computer simulation data for the compressibility factor; **(ii)** accuracy is preferred to simplicity; **(iii)** the radius of convergence must be $b>0.64$.

**(i)** *Consistency.* There are two ways to check the consistency between the calculated accurate virial coefficients and the computer simulation data for the compressibility factor. The first is that, if we include more accurate virial coefficients in the virial equations of state, the EOS should approach the computer simulation data more accurately. The question then is what is the appropriate order for the virial equation to reproduce the computer simulation data. To answer this question, we considered the virial EOS, Eq. (1), using the virial coefficients given by Clisby and McCoy.[9-10] We thus generated ten virial EOSs, $Z_J$ with $J = 1$ to 10. Then we compared the $Z$ values with the computer simulation data given by Wu and Sadus[27] in the stable density range from 0.04 to 0.95 in reduced units (32 data points), with the Kolafa *et al.*[23] data in the density range from 0.20 to 1.03 (31 data points) which includes the metastable region ($\rho > 0.943$), and finally with the Kolafa *et al.* data[23] but only for the metastable region from 0.95 to 1.03 (9 data points). Table 1 lists the absolute average deviations (AAD, %) for each virial EOS. Figure 1 shows the comparison for a selection of these virial EOSs.

One observes in the table that only when more than 5 virial coefficients are considered can the data in the stable region be reproduced with an AAD below 5%. In particular, as can be seen in Fig. 1, the $Z_5$ EOS can reproduce the data only for $y \leq 0.3$ (percentage deviations below 4% with respect to computer simulation data), and $Z_7$ up to $y$ around 0.4 ($\rho < 0.8$) (percentage deviations below 3% for every data in this region). When nine virial coefficients are used in Eq. (1), all the Wu and Sadus computer simulation data in the stable region can be reproduced with individual deviations below 4%, the overall AAD being below 1%. With $Z_{10}$ the results are very similar.

When the Kolafa et al. data[23] are used as reference, including both stable and metastable densities, the AADs obtained (AAD2 in Table 1) are obviously higher than when only the stable region is considered. Nevertheless, one observes in Table 1 that for $Z_9$ and $Z_{10}$ the AADs are practically the same regardless of whether only the stable region (AAD1) or the full range (AAD2) is considered. Finally, if only the metastable region is considered, the lowest AAD is 3.4% with 10 virial coefficients (the individual deviations being below 5%).

As can be seen in Fig. 1 (in which $Z_{10}$ is not plotted because it is practically the same curve as $Z_9$ at the scale of the graph) and Table 1, the virial equation to 9th order can be considered as adequate for the stable region and moderately adequate for the metastable region (the individual deviations being less than or equal to 7%). In the case of the $Z_{10}$ EOS, the individual deviations are below 5%. The more accurate are the virial coefficients used in Eq. (1), the more accurate the derived $Z$ results.

The other way to check the consistency is that the computer simulation data should yield correct virial coefficients. Because the published computer simulation data for a single method does not reach the sufficient detail, we could not do this check in the present work.

**(ii)** *Accuracy is preferred to simplicity.* In Eq. (9) the number of variables is $(j-i+1)$ if $i>0$, and $(j-i+2)$ if $i<0$. (Note that $j \geq 1$, see the bottom of point (iii) below). Expanding Eq. (9) and setting each first virial coefficient equal to the first accurate values, one can obtain the variables $a_k$ and $b$. Because only the first ten virial coefficients are calculated accurately, the number of variables ranges from 1 to 10. If we take it to be 10, then the first ten accurate virial coefficients are obtained, but the resulting EOSs cannot be as simple as the CS expression.

**(iii)** *The radius of convergence must be b>0.64.* For the face-centred cubic lattice,[28] $b_{fcc} = \pi/\sqrt{18} \approx 0.7405$. For random close packing in three dimensions,[29-30] the common value of $b$ is conjectured to be $b_{rcp} \approx 0.64$. For hard sphere fluids, the close packed-value is expected to be in the equation of state, then Eq. (9) must been used with $j \geq 1$. In any case, when the HS EOSs is going to be used together a perturbation attractive term, some the value of *b* can be properly chosen in order to adequately represent different properties, as are the phase diagram[22,31].

In accordance with the above three constraints, we obtained 57 possible EOSs in the form of Eq. (9). For 56 of them, $j \in [1, 7]$ and $i$ may be one of several positive or negative integers. The last EOS is that obtained with $i = 0, j = 8$. From a check of these 57 EOSs, we chose that with $i = -5, j = 2$:

$$Z_{AEM} = \sum_{k=-2}^{5} a_k (y - b_1)^k \qquad (10)$$

where $b_1 = 0.9262135992$, and parameters $a_k$ are given in Table 2.

The other options were rejected because negative virial coefficients for $B_{15}$ appeared, the eleventh virial coefficient was too large compared with the value 127.93 predicted by Clisby and McCoy[10], or $b < 0.64$.

**Results**

From Fig. 1 and Table 1, one observes that a virial equation with the first nine and ten accurate virial coefficients can describe the stable region and the low density metastable region well, but fails moderately in the very high metastable region. More accurate virial coefficients are required to study the order at which the virial equation can describe the whole region more accurately. As of now, one knows that the tenth order is insufficient.

In the present work, we propose Eq. (10) to adequately reproduce at least the first nine virial coefficients and the compressibility factor values over the whole range, and we are interested in knowing whether higher coefficients can be predicted and how accurately it reproduces the computer simulation data for the compressibility factor when compared with other commonly used or recently proposed EOSs.

Table 3 lists the results for the first sixteen virial coefficients from Eq. (10) and compares them with the accurate data of Clisby and McCoy[9-10] for the first ten virial coefficients and also with the estimated values for the higher ones. We compare also with the values predicted by other EOSs, in particular, the CS[15] EOS, as example of a very simple expression, the CM1 [10], CM2 [10], KLM1[23], KLM2[23], and Liu[16] EOSs, as very accurate and more complex recently proposed expressions, and the relatively simple and more recently proposed SH[26] EOSs. For virial coefficients higher than the fourth one, the percent uncertainty of the Clisby and McCoy data[10] are given. For $B_5$ to $B_9$, the percent deviations of the values predicted by the EOSs and the above mentioned ones are given only if they are clearly greater than those uncertainties. For $B_{10}$ and higher virial coefficients, all the percent deviations are given.

As is well known, the Carnahan-Starling equation gives integer values for the virial coefficients. In any case, Table 3 shows that it gives at least qualitatively similar results for most of the coefficients considered here, which can be considered as a certain success in view of its simplicity. Moreover, it gives values inside the uncertainty for the estimated values of $B_{13}$, $B_{14}$ and $B_{16}$, giving better results than some other more complex EOSs.

The CM1 and CM2 EOSs were constructed by taking into account the virial coefficient data proposed by Clisby and McCoy, and the results in the table show that they reproduce the first ten values accurately. Nevertheless, the deviations increase as the order of the coefficients increases. The best results are obtained by using CM1, which give accurate results (inside the uncertainty deviation of the reference data) for first eleven virial coefficients and also for $B_{14}$.

It is interesting to know whether some other more recent EOSs can reproduce the higher predicted virial coefficients better. In particular, the KLM1 and KLM2 EOSs cannot reproduce the seventh and higher virial coefficients inside the data uncertainty, with the exception of $B_8$, $B_9$, and the estimated $B_{14}$ when KLM1 is used. In general, these EOSs give higher percent deviations than the simple CS equation for most of the coefficients higher than the tenth. In particular, we would note that the estimated value for $B_{16}$ from KLM1 is excessively low, whereas KLM2 value is quite high.

As can be seen in Table 3, the Liu EOS can only reproduce exactly the second virial coefficient, deviates less than a 1.4% from the accurate values for the first nine coefficients, but gives clearly greater deviations for the rest of the coefficients.

The SH EOS is a simple expression, with a lower number of parameters, and constructed by using the first seven virial coefficients, but for higher ones the obtained percentage deviations are higher than the inaccuracies in the data. In particular, the relative errors of virial coefficients $B_{10} \sim B_{16}$ are generally higher than the obtained with CM1 and CM2 EOS, which are analytically more complex.

Finally, our proposal, Eq. (10), is constructed by using the first nine virial coefficient, and then it is the one giving the lowest deviations for a largest number of coefficients. For virial coefficients higher than the ninth one, the obtained percentage deviations are lower than the uncertainties of the estimated Clisby and McCoy values[9-10].

In order to test the accuracy of Eq. (10) in reproducing the compressibility factor values given by accurate computer simulation data, we considered first the stable and metastable ranges separately, and then together.

For the stable range, we took as reference the data of Wu and Sadus[27] in the density range from 0.04 to 0.95, and for the metastable region the Kolafa *et al.* [23] data for densities from 0.95 to 1.03 together with the Kolafa[32] data from 1.02 to 1.09.

The results for the AADs between EOSs and computer simulation data in each range are given in Table 4. When only the stable range is considered (AAD1), the highest deviation was for CS EOSs. All the others, including our new proposal, give very similar results.

When both stable and metastable regions are considered together (AAD2), the KLM1 and KLM2 give clearly the best results. Obviously, this is due to the fact that some of the coefficients in those EOSs (in particular, three coefficients for KLM1 and four for KLM2) were obtained from a fitting procedure to the same data considered here. The CM1 and CM2 EOSs contain 9 fixed coefficients, and in this range give practically the same result as our new Eq. (10) which also contains 9 coefficients, although our proposal gives better predictions for high virial coefficients.

When only the metastable region is considered (AAD3), the CM1 EOS gives the lowest AAD value, but only with a minor improvement over KLM1 and our new proposal.

A shortcoming of Eq. (10) is that its $b_1$ value is larger than the random closed packed-value 0.64 for hard sphere fluids. In order to solve this problem, we substituted $b = 0.64$ into Eq. (10) and repeated the calculations with accurate virials up to $B_{10}$ and the same constraints on $i$ and $j$ as before. There are now 56 equations involved, but none of them is comparable with Eq. (10) except for the $b$ value. The results strongly depend on the higher virial coefficients. If more accurate virials become known, work in this line will be interesting in the future.

**Conclusions**

In this paper, we have extended the asymptotic expansion method proposed by Khanpour and Parsafar in considering accurate values for the first ten virial coefficients. First, we checked the convergence of the virial EOSs, and found that the first nine and ten accurate virial coefficients can describe the stable and the low density metastable regions well, but fail moderately in the very high metastable region.

A new EOS was then proposed by choosing from among 57 possible analytical expressions. The results for the virial coefficients and compressibility factor were compared against other well-known or recently proposed EOSs and also against accurate data from computer simulations and calculations, including estimated accepted values for the virial coefficients higher to the tenth. The resulting proposal, Eq. (10), is the only one published to date giving very low deviations for the accurate values for the first ten virial coefficients and estimated values for the higher ones. coefficients. Moreover, it accurately reproduces the compressibility factor values from computer simulations in both the stable and the metastable ranges. Further development of this work in greater depth will need to be along two lines of inquiry: the calculation of accurate higher order virial coefficients, and accurate computer simulation data of $Z$ versus $\rho$.

**Acknowledgement**

The National Natural Science Foundation of China under Grant No. 10804061, the Natural Science Foundation of Shandong Province under Grant No. Y2006A06, and the foundation from QFNU and DUT have supported this work (J.T., H.J., and Y.G.). A.M. thanks the Ministerio de Educación y Ciencia of Spain for support through Project FIS2006-02794 FEDER.

**Notes and references**

1. A. Mulero, *Theory and Simulation of Hard-Sphere Fluids and Related Systems*, Lect. Notes Phys. 753 (Springer, Berlin Heidelberg, 2008).

2. L. V. Yelash, T. Kraska, E. A. Muller, and N. F. Carnahan, Phys. Chem. Chem. Phys., 1999, **1**, 4919.

3. P. Paricaud, A. Galindo, and G. Jackson, Fluid Phase Equilibr., 2002, **87**, 194.

4. A. Mulero, C.A. Galán, M.I. Parra, and F. Cuadros. Lect. Notes Phys., 2008, **753**, 37.

5. A. Malijevsky and J. Kolafa, Lect. Notes Phys., 2008, **753**, 27.

6. B. R. A. Nijboer, L. van Hove, Phys. Rev., 1952, **85**, 777. Also see, Ref. (1), Chapter 2.

7. S. Labik, J. Kolafa, A. Malijevsky, Phys. Rev. E, 2005, **71**, 021105.

8. R.D. Rohrmann, M. Robles, M. López de Haro and A. Santos, J. Chem. Phys., 2008, **129**, 014510.

9. N. Clisby and B. M. McCoy, Pramana-J. Phys., 2005, **4**, 3609.

10. N. Clisby and B. M. McCoy, J. Stat. Phys., 2006, **122**, 15.

11. M. S. Wertheim, Phys. Rev. Lett., 1963, **10**, 321.

12. E. Thiele, J. Chem. Phys., 1963, **39**, 474.

13. A. I. Rusanov, J. Chem. Phys., 2004, **121**, 1873.

14. A. Malijevsky, J. Veverka, Phys. Chem. Chem. Phys., 1999, **1**, 4267.

15. N. F. Carnahan and K. E. Starling, J. Chem. Phys., 1969, **51**, 635.

16. H. Liu, arXiv:cond-matt/0605392.

17. B. Barboy, W.M. Gelbart, J. Chem. Phys., 1979, **71**, 3053.

18. M. Baus, J.L. Colot, Phys. Rev. A, 1987, **36**, 3912.

19. I.C. Sanchez, J. Chem. Phys., 1994, **101**, 7003.

20. W. Wang, M.K. Khoshkbarchi, J.H. Vera, Fluid Phase Equil., 1996, 115, 25.

21. R.J. Speedy, J. Phys.: Cond. Matt., 1997, **9**, 8591.

22. L.V. Yelash and T. Kraska, Phys. Chem. Chem. Phys., 2001, **3**, 3114.

23. J. Kolafa, S. Labʹık, A. Malijevskʹy, Phys. Chem. Chem. Phys., 2004, **6**, 2335.

24. M. Khanpour and G.A. Parsafar, Chem. Phys., 2007, **333**, 208.

25. M. Khanpour and G.A. Parsafar, Fluid Phase Equilibr., 2007, **262**, 157.

26. A. Santos and M. López de Haro, J. Chem. Phys., 2009, **130**, 214104.

27. G. W. Wu and R. J. Sadus, AICHE J., 2005, **51**, 309.

28. T. C. Hales, Ann. Math., 2005, **162**, 1065.

29. R. D. Kamien and A. J. Liu, Phys. Rev. Lett., 2007, **99**, 155501.

30. A. V. Anikeenko and N. N. Medvedev, Phys. Rev. Lett., 2007 **98**, 235504.

31. L. V. Yelash, T. Kraska and U. K. Deiters, J. Chem. Phys., 1999, **110**, 3079.

32. J. Kolafa, Phys. Chem. Chem. Phys., 2006, **8**, 464.

**Figure 1**. The curve of Z versus $y$. Points are data from computer simulations by Wu and Sadus[27] and Kolafa et al.[23] The Kolafa et al. data at lower densities are not shown because they are practically identical to the Wu and Sadus case. The lines are the virial equations, Eq. (1), with accurate virial coefficients.

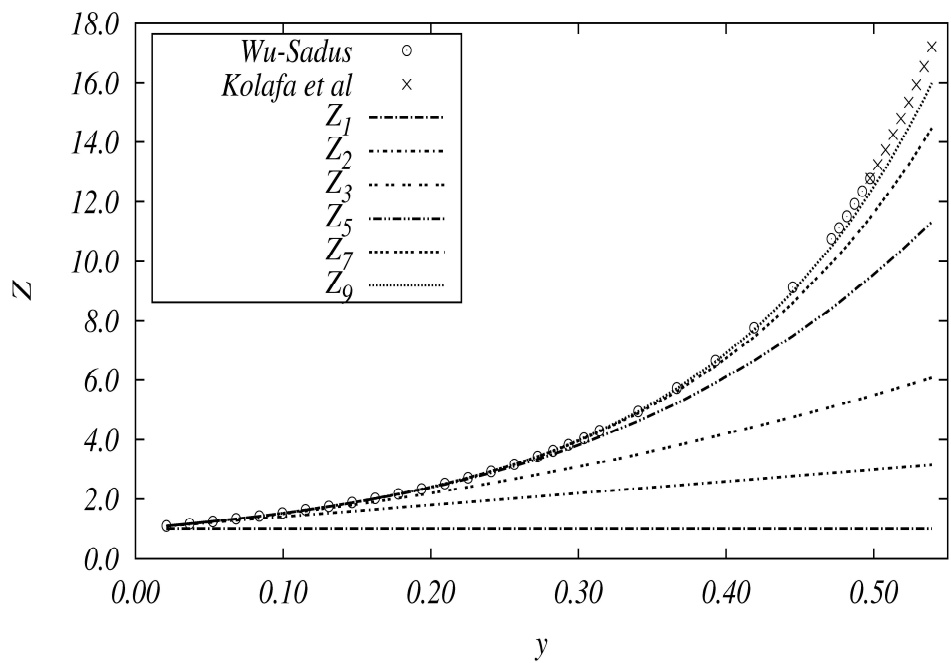

**Table 1**. AAD (%) of virial equations with accurate virial coefficients compared with the computer simulation data. VE: virial equation; AAD1: the comparison with simulation data by Wu and Sadus[27] in the stable range from $\rho = 0.04$ to $\rho = 0.95$, i.e., $y \in [0.02, 0.50]$, 32 data points; AAD2: the comparison with simulation data by Kolafa *et al.*[23] in the range from $\rho = 0.20$ to $\rho = 1.03$, $y \in [0.10, 0.54]$, 31 data points; AAD3: comparison with data by Kolafa *et al.*[23] in the metastable range from $\rho$ = 0.95 to 1.03, $y \in [0.50, 0.54]$.

|  | $z=1$ | $z_{J=2}$ | $z_{J=3}$ | $z_{J=4}$ | $z_{J=5}$ | $z_{J=6}$ | $z_{J=7}$ | $z_{J=8}$ | $z_{J=9}$ | $z_{J=10}$ |
|---|---|---|---|---|---|---|---|---|---|---|
| **AAD1 (%)** | 63.15 | 38.29 | 22.22 | 12.65 | 07.20 | 04.11 | 02.34 | 01.39 | 00.86 | 00.55 |
| **AAD2 (%)** | 81.00 | 59.43 | 40.34 | 26.09 | 16.46 | 10.21 | 06.25 | 03.79 | 00.84 | 00.51 |
| **AAD3 (%)** | 93.22 | 79.18 | 61.03 | 43.77 | 30.02 | 19.97 | 12.98 | 08.32 | 05.29 | 03.35 |

**Table 2**. Coefficients $a_k$ for the new HS EOSs, Eq. (10).

| $k$ | -2 | -1 | 0 | 1 |
|---|---|---|---|---|
| $a_k$ | 5.489785755 | 10.29617715 | 8.100015583 | 2.394846562 |
| $k$ | 2 | 3 | 4 | 5 |
| $a_k$ | -1.419388208 | -2.165373211 | -1.097171967 | -0.2050878768 |

**Table 3**. The virial coefficients predicted by equations of state. * represents the predictive values in Ref. (9-10). Contents in parenthesis represent the relative errors.

|  | Refs. (9-10) | CS | CM1 | CM2 | KLM1 | KLM2 | Liu | SH | Eq. (10) |
|---|---|---|---|---|---|---|---|---|---|
| $B_2$ | 4 | 4 | 4 | 4 | 4 | 4 | 4 | 4.000003 | 4 |
| $B_3$ | 10 | 10 | 10 | 10 | 10 | 10 | 10.021445 | 10.000003 | 10 |
| $B_4$ | 18.364768 | 18 | 18.364768 | 18.364768 | 18.364768 | 18.364768 | 18.216470 | 18.364769 | 18.364768 |
| $B_5$ | 28.224512 (0.9 $10^{-3}$ %) | 28 (0.8%) | 28.224511 | 28.224510 | 28.224450 | 28.224450 | 28.357348 (0.5%) | 28.224504 | 28.224512 |
| $B_6$ | 39.815148 (2.3 $10^{-3}$ %) | 40 (0.46%) | 39.815146 | 39.815146 | 39.815470 | 39.815470 | 40.288163 (1.2%) | 39.815125 | 39.815148 |
| $B_7$ | 53.344420 (7 $10^{-3}$ %) | 54 (1.2%) | 53.344456 | 53.344455 | 53.270025 (0.14%) | 53.385486 (0.08 %) | 53.811465 (0.88%) | 53.344379 | 53.344420 |
| $B_8$ | 68.537549 (2.6 $10^{-2}$ %) | 70 (2.1%) | 68.538722 | 68.538721 | 68.541201 | 68.735949 (0.29%) | 68.691231 (0.22%) | 68.608510 (0.10%) | 68.537549 |
| $B_9$ | 85.812838 (0.1%) | 88 (2.6%) | 85.818013 | 85.818015 | 85.868942 | 85.402287 (0.5%) | 84.666099 (1.3%) | 85.531747 (0.33%) | 85.812838 |
| $B_{10}$ | 105.775104 (0.37%) | 108 (2.10%) | 105.731518 (0.04%) | 105.731519 (0.04%) | 106.192058 (0.39%) | 103.471699 (2.18%) | 101.504253 (4.04%) | 104.320185 (1.38%) | 105.405615 (0.35%) |
| $B_{11}*$ | 127.93 (0.82%) | 130 (1.62%) | 128.37 (0.34%) | 126.75 (0.92%) | 130.91 (2.33%) | 124.81 (2.44%) | 119.11 (6.89%) | 124.86 (2.40%) | 127.58 (0.27%) |
| $B_{12}*$ | 152.67 (0.28%) | 154 (0.87%) | 154.27 (1.05%) | 149.83 (1.86%) | 160.41 (5.07%) | 154.02 (0.88%) | 137.71 (9.80%) | 147.25 (3.55%) | 152.61 (0.04%) |
| $B_{13}*$ | 181.19 (0.93%) | 180 (0.66%) | 184.22 (1.67%) | 177.40 (2.09%) | 190.82 (5.31%) | 197.74 (9.13%) | 158.18 (12.70%) | 171.28 (5.47%) | 180.82 (0.20%) |
| $B_{14}*$ | 214.75 (3.1%) | 208 (3.14%) | 218.69 (1.83%) | 203.23 (5.36%) | 208.86 (2.74%) | 260.05 (21.09%) | 182.48 (15.03%) | 197.26 (8.14%) | 212.56 (1.02%) |
| $B_{15}*$ | 246.96 (1.1%) | 238 (3.63%) | 258.44 (4.65%) | 229.40 (7.11%) | 185.45 (24.91%) | 330.61 (33.87%) | 214.52 (13.14%) | 224.98 (8.90%) | 248.21 (0.51%) |
| $B_{16}*$ | 279.17 (3.9%) | 270 (3.28%) | 304.58 (9.10%) | 267.73 (4.10%) | 70.43 (74.77%) | 361.62 (29.53%) | 261.30 (6.40%) | 254.25 (8.93%) | 288.19 (3.23%) |

**Table 4**. AAD (%) for $Z$ from EOSs compared with the computer simulation data. AAD1 and AAD2 have the same meaning as in Table 1. AAD3: the comparison with simulation data by Kolafa *et al.*[23] and Kolafa [32] in the metastable range from $\rho = 0.95$ to $\rho = 1.09$. Data from from $\rho = 0.95$ to $\rho = 1.01$ are from Kolafa *et al.*[21], and data from $\rho = 1.02$ to $\rho = 1.09$ are from Kolafa [32].

| AAD (%) | CS | CM1 | CM2 | KLM1 | KLM2 | Liu | SH | Eq. (11) |
|---|---|---|---|---|---|---|---|---|
| AAD1 | 0.22 | 0.18 | 0.16 | 0.16 | 0.16 | 0.17 | 0.17 | 0.17 |
| AAD2 | 0.21 | 0.06 | 0.08 | 0.02 | 0.001 | 0.19 | 0.12 | 0.06 |
| AAD3 | 0.74 | 0.59 | 0.79 | 0.64 | 1.23 | 2.91 | 0.91 | 0.67 |